# Magnification of mantle resonance as a cause of tectonics

Mensur Omerbashich

*Physics Department, Faculty of Science, University of Sarajevo, Zmaja od Bosne 33, Sarajevo, Bosnia*
*Phone +387-63-817-925, Fax +387-33-649-359, E-mail: momerbasic@pmf.unsa.ba; cc: omerbashich@gmail.com*

**Abstract.** Variance-spectral analysis of superconducting gravimeter (SG) decadal data (noise inclusive) is presented suggesting that the Earth tectonogenesis is based on magnification of the mass (mainly the mantle) mechanical resonance, in addition to or instead of previously hypothesized causes. Here, the use of raw (gapped and unaltered) data is regarded as *the* criterion for a physical result's validity, so data were not altered in any way. Then analogously to the atmospheric tidal forcing of global high-frequency free oscillation, I propose that the Moon's synodically recurring pull could likewise drive the long-periodic (12-120 minutes) oscillation of the Earth. To demonstrate this, I show that the daily magnitudes of mass (gravity) oscillation, as a relative measure of the non-stationary-Earth's kinetic energy, get synodically periodic while correlating up to 0.97 with seismic energies on the day of shallow and 3 days before deep earthquakes. The forced oscillator equations for the mantle's usual viscosity and the Earth springtide and grave mode periods successfully model an identical 3-days phase. Finally, whereas reports on gravest earthquakes (of ~M9.5) put the maximum co-seismic displacement at ~10 m, the same equations predict the maximum displacement as 9.8 m, too. Hence, the same mechanism that causes bridges to collapse under the soldiers step marching could be making the lithosphere fail under the springtide-induced magnification of mantle's resonance resulting in strong earthquakes of unspecified type (most of some 400 earthquakes that affected the SG in 1990-ies were strike-slip and thrust). If this assertion is correct, then many if not most large earthquakes could be predictable – spatially and temporally – by monitoring the tidal v. grave mode oscillation periodicities of separate mass bodies in the upper mantle and in the crust.

## 1. Gravimetric Terrestrial Spectroscopy

I use superconducting gravimeter (SG) 1Hz observations to show that it could be useful to regard the Earth a forced mechanical oscillator in which resonant magnification of Earth *total-mass* (i.e., atmosphere to inner core, but mainly mantle) resonance occurs. This means that the Earth tectonics and related phenomena could be also mechanically caused, in addition to or instead of thermo-nuclear-chemical causes hypothesized in the past. The lunar synodic semi-monthly forcing drives the long periodic (here between 12-120 minutes) oscillation magnitudes of the Earth's total mass (geophysical noise inclusive).

Twenty-some superconducting gravimeters (SG) are used worldwide for studying the Earth tides, the Earth rotation, interior, ocean and atmospheric loading, and for verifying the Newtonian constant of gravitation [1]. I used gravity data from the Canadian SG at Cantley, Quebec. This stable [2] instrument is sensitive to about one part in $10^{12}$ of surface gravity at tidal and normal mode frequencies, so it records antipodean earthquakes as small as M~5.5 [3]. To process the SG-gravity gapped records of 1Hz output I constructed a non-equispaced filter with 2-sigma Gaussian, and 8-seconds filtering step as recommended by *Global Geodynamics Project*.

Since the masses of gas, water, rock, mantle, and core together account for Earth gravity, the state and perturbations of all the Earth masses, including atmo-sphere, ocean, land, and interior are of interest here. I there-fore regard the entire information as my signal, which is then composed of classical geophysical signal plus classical geophysical noise. Useful-ness of geophysical noise for studying the Earth interior has been demonstrated [4]. Of all SG locations, the Canadian was the best for this purpose as it is antipodal to the seismically most active region on Earth – the Pacific Rim. This ensured that the signal strength was max-imal. Then the whole (total-mass) Earth can be studied by means of raw gravity observations (further: gravity observations) that are neither stripped of tides nor corrected for environ-mental effects. This can be achieved by looking into the Earth gravity spectra from the band of Earth long eigenperiods 12–120 min, or low eigenfrequencies 12–120 cycles per day (c.p.d.). I term my approach the *Gravimetric Terres-trial Spectroscopy* (GTS).

Therefore in the following, I regard the use of raw (gapped and unaltered) data as *the* criterion for a physical result's validity [5]. Thus no raw data were altered in any way for all the essential compu-tations; however, the raw data were treated classically for all testing purposes. Data were obtained in a Government of Canada (a non-public domain) release rest-ricting distribution to third parties. Most of data processing presented in this paper was completed before November 2003, and all the computations by the end of 2005.

## 2. Methodology

Gravity spectra were obtained in the Gauss-Vaníček spectral analysis (GVSA) of over ten billion non-equispaced, Gaussian-filtered SG gravity decadal recordings from the 1990-ies. GVSA fits, in the least-squares sense, data to trigonometric functions. GVSA was first proposed by [6], was first developed by Vaníček [7] [8], and was simplified by others [9] [10]. Magnitudes of GVSA-derived, variance-spectrum peaks depict the contribution of a period to the variance of the time-series, of the order of (some) % [7]. Thus, variance spectra, expressed in var%, or power spectra, expressed in dB [11] can be produced from incomplete numerical records of any length. Thanks to its many important advantages, GVSA was a more suitable technique for GTS than any of the typically used tools such as the Fourier spectral analysis [5] [12]. GVSA seemed most apt for GTS primarily due to its: (i) ability to handle gaps in data[13], (ii) straight-forward significance level regime[14], and (iii) distinctive and virtually unambiguous depiction of background noise levels[3]. Just like the variance is the most natural description of noise in a record of physical data, a variance spectrum tells naturally and simultaneously of the strength of cyclic signal's imprints in noise and thereby of signal's reliability too [5]. These advantages make GVSA a unique field descriptor that can accurately and simultaneously estimate both the structural eigenfrequencies and field relative dynamics [5]. Over the past thirty years, GVSA was applied in astronomy, geophysics, medicine, microbiology, finances, climatology, etc. See [3] for more GVSA references and a blind performance test using synthetic data.

The 2-sigma Gaussian was selected for filtering the SG gravity data in a non-equispaced fashion, meaning that the sub-step gaps were accounted for by Boolean-weighting each measurement. This filter's response stayed well above 90 var% over the entire band of interest, passing all the systematic contents from 1 min to 10 years. At no point in this research did any of the gravity spectra' oscillation magnitudes exceed 1 var%, and most of the time they stayed the safe two orders of magnitude beneath that level – in the order of 0.01 var%. The equispaced Gaussian weighting function $w$, used here for filtering of series of step size $\Delta t$ with $(2N+1)$ elements, for $n$ observations $l(t)$ at the time instant $t$, and for selected $\sigma = 2$, is [12]:

$$w_i(\Delta t, \sigma) = \frac{1}{2\sigma\sqrt{2\pi}} e^{\frac{(i\Delta t)^2}{2\sigma^2}}, \quad (1)$$
$$\forall i = -N, -N+1, \ldots, 0, 1, \ldots, N$$

and the filter [12]:

$$l_j^*(l, w) = \frac{1}{\sum w} \cdot \sum_{i=0}^{N} l_{j+i}(t) \cdot w_i(\Delta t), \quad (2)$$
$$\forall\; j = 1, \ldots, n$$

becoming a Boxcar case for $w_i = 1/(2N+1)\; \forall\; i \in \aleph$. The guiding idea behind the non-equispaced filter was to enable rigorous data processing in which no low-frequency information would be lost due to filtering, unlike in equispaced filters (usually applied for Fourier methods), where variation in the original ratio of populated $p_i = \{l_1, l_2 \ldots\}$ versus empty place-holders $q_i = \{\}$ is overlooked. Thus data distortion by contrivance of invented values that must fall on the integer number of steps takes place in cases when equispaced filters are used. When a portion of the record lacks observations, its average should be renormalized regardless of the choice of the filtering function, as:

$$l_i^{**} = l_i^* \Big/ \sum_{\substack{\text{existing} \\ \text{data pts.}}} l_i^* \; : \; \sum_{\substack{\text{existing} \\ \text{data pts.}}} l_i^{**} = 1, \quad (3)$$

where $l_i^{**}$ are re-normalized filtered values. See [3] for non-equispaced filter implementation code, as well as the list of earthquakes used.

Given that SG measurements are used here to study the whole Earth, and based on the fact that it is the original and irregular impulses which affect the instruments where the superimposed vibrations probably pass unnoticed [15], in what follows I do a superimposed epoch analysis. Thus based on Jeffreys's rule of thumb ("*In many earthquakes observations of only the horizontal Earth movements during the passage of shear (S) waves can be used to estimate the order of the total released energy.*" [16], and the Earth thought of as a simple mechanical oscillator [17] and taken as a viscoelastic continuously vibrating stopped up mechanical system [18], assume valid the following

> **Physical hypothesis "A":**
> *The ratio of seismic energy $E_S$ and total kinetic energy $E_K$ on Earth is constant.*

Note that most earthquakes used here were tectonic thrust and strike-slip, and that "total" in the above refers to the non-stationary Earth with negligible non-impulse $E_K$. Seismic energy, $E_S$, as that part of the total kinetic energy transmitted by the lithosphere is normally found from earthquake magnitudes estimated at seismic observatories worldwide. Since earthquakes are almost exclusive source of kinetic energy, and the lithosphere makes merely ~1/50 of the Earth's volume and ~1/100 of the Earth's mass [19], the hypothesis "A" generally holds. Seismic energy expressed in units of ergs is computed using modified Richter-Gutenberg empirical formula [20]:

$$\log_{10} E_S = 1.5 \cdot M_S + 11.4 . \quad (4)$$

Physically, magnitudes of the gravity field oscillations are proportional to the kinetic energy, $E_K$, needed to displace the Earth's *inner masses* (the Earth minus atmosphere and lithosphere) as the medium, from the state of rest to that of unrest [21]. I therefore computed as a relative measure of the Earth kinetic energy the series of simple-average magnitudes $^T\bar{\mu}_E$ of the Earth oscillation at particular normal frequency $\omega$, as determined from a gravity record spanning a specific time interval $\Delta T$ (<u>W</u>eek, <u>D</u>ay, <u>H</u>our, etc.). Then for some normal mode of Earth oscillation the amplitude $^T\bar{\mu}_E$ of the gravity spectrum $s^{GVSA}(\omega)$ is computed as average of three (minimum spectrum size in GVSA):

$$^T\bar{\mu}_\omega = \frac{1}{3}\sum_{i=0}^{2} s(\omega_i) , \quad (5)$$

$$\omega_i = \begin{cases} 1/[(\text{mode period})^{[\text{sec}]} - 1^{[\text{sec}]}], & \text{for } i = 0 \\ 1/[(\text{mode period})^{[\text{sec}]}] & , \text{for } i = 1 \\ 1/[(\text{mode period})^{[\text{sec}]} + 1^{[\text{sec}]}], & \text{for } i = 2 \end{cases}$$

As a test, Fig. 1 depicts the average oscillation magnitudes $^T\bar{\mu}_E$ (colored lines) over all (here 1000) spectral points in the band of interest $\mathbf{E}_l \in$ 12–120 c.p.d. from 2, 3, and 4 weeks of gravity data past a great earthquake. Earth ringing is thus observed and measured relatively by means of SG gravity variance-spectra for at least six weeks past an earthquake of M7.5 or stronger; see Fig. 1. This is in agreement with the solid-tide general dissipation rate of 83±45GW [22].

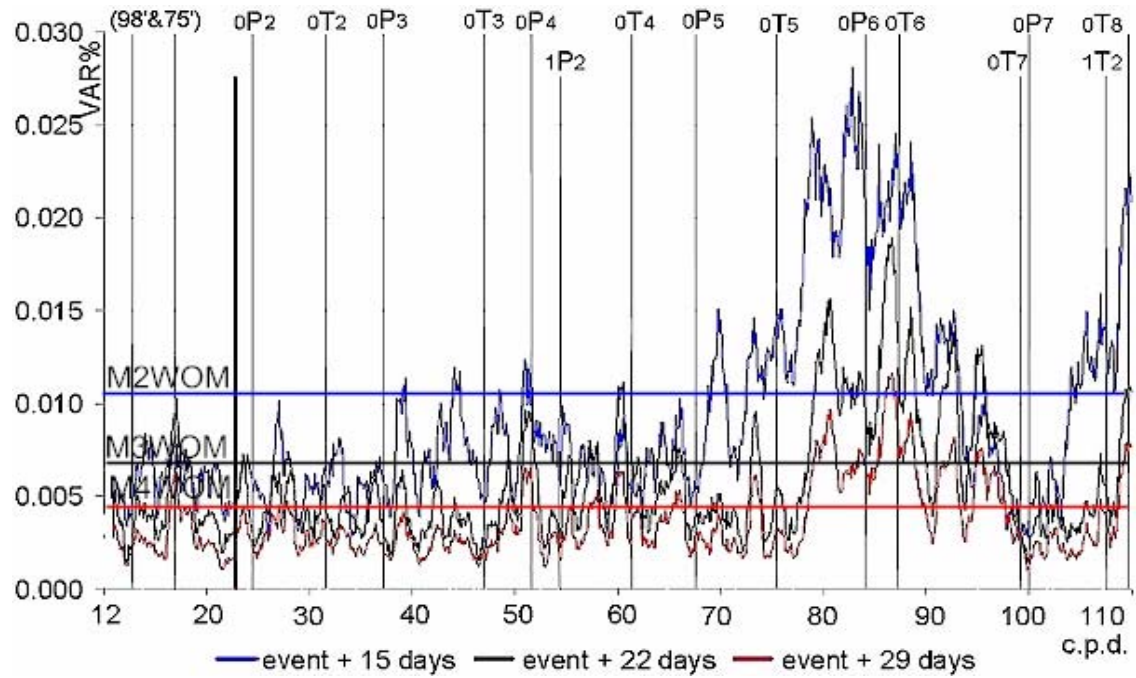

**Fig.1.** Gravity variance-spectra as a relative measure of kinetic energy. <u>M</u>ean-<u>W</u>eekly <u>O</u>scillation <u>M</u>agnitudes (MWOM), from variance spectra of detided gravity at Cantley from 2 (blue), 3 (black) and 4 (red) weeks of data past the M8.8 Ballenys Islands earthquake [68], of 25 March 1998, Harvard CMT-032598B. Focus was in Tasman fracture zone between the Southeast Indian and Pacific-Antarctic ridges, 700 km east of South magnetic pole along winter ice boundary and 600 km north of George V Coast. East Antarctica is a stable Precambrian shield composed of 3+ billion years old metamorphic rocks that did not undergo major change in recent geological times. Normal periods from Zharkov model. Filter step 32 s. Resolution 1000 pt.

## 3. Gravity-seismicity correlation

Based on the starting physical hypothesis "A" linear correlation is sought between the energy in Earth's gravitational normal modes (series $X_{\bar{\mu}}$ of magnitudes of gravity daily oscillation from the 1990-ies), and the energy released by earthquakes (the series $Y_S$ of seismic energies and for check of seismic magnitudes from 381 medium-to-large global earthquakes found (by visual inspection) to have excited the Canadian SG record spanning the last decade). Cross-correlation functional values were computed between the $n_\omega$ compo-nents of vector $X_{\bar{\mu}}$ and the vector of earthquake energies, $Y_S$, as [23]:

$$\Omega(X,Y;u) = \frac{\sum_{i=0}^{30}\left\{\left[X(\tau_0 + i\Delta\tau) - \overline{X}\right]\left[Y(\tau_0 + (i+j)\Delta\tau) - \overline{Y}\right]\right\}}{\sqrt{\sum_{i=0}^{30}\left[X(\tau_0 + i\Delta\tau) - \overline{X}\right]^2 \sum_{i=0}^{30}\left[Y(\tau_0 + (i+j)\Delta\tau) - \overline{Y}\right]^2}}, -1 \leq \Omega_{x,y} \leq 1, \qquad (6)$$

where i = 1, 2... $n_\omega$; $u = j\cdot\Delta\tau \wedge j \in \mathbb{N}$, $\Delta\tau$ = 1 day. Here $\Omega_{x,y}(u)$ is the value of the correlation function $\Omega$ between the components of $X_{\bar{\mu}}$ and $Y_S$ lagged by $u$ days (here from | $u$ | = 0, 1 ... 30 days).

Seismically induced high-frequency Earth oscillation radiates energy in the order of megawatts [18], versus low-frequency (solid-Earth and oceanic) tidal dissipation that amounts to about 2TW [24]. Also, the Earth's incessant high-frequency oscillation, beating in high eigenfrequencies down to 2.2 mHz or at short eigenperiods up to ~7½ minutes [18], radiate each day an amount of energy normally released by a single M5.75–M6 earthquake [25]. For these reasons the magnitudes of Earth high-frequency oscillation are not of interest here. Besides, if high correlations could be obtained without using magnitudes of high-frequency Earth oscillation, then this setup will physically satisfy for the entire natural band as well. Note that [26] and [27] first offered an explanation for the "incessant" short-periodic Earth oscillation, while [25] offered a more comprehensive explanation of that phenomenon.

The choice of surface magnitudes in Eq. (4) rather than moment magnitudes was not just preferential. Namely, more than 96% of all >M6.39 earthquakes that affected the Cantley SG record were wea-ker than M7.5 (when the scales based on surface magnitudes start saturating). Hence, using surface magnitudes rather than moment magnitudes represents a more stringent approach given that the linear correlation Eq. (6) can be sensitive to a small number of relatively large input values. If high correlations could be obtained using the surface magnitudes, then this setup will satisfy for moment magnitudes as well.

Pairs ($X_{\bar{\mu}}$, $Y_S$) of physically dependent values are not random samples from *bivariate normal distribution* [28], so that the confidence intervals for correlation coefficients cannot be computed [29]. $X_{\bar{\mu}}$ values obtained always from a large sample of up to 86 400 normally distributed gravity measurements per day are not normally distributed either, instead those values follow the $\beta$–distribution [30]. In addition, seismic magnitudes used in vector $Y_S$ follow the Boltzmann distribution [31]. Hence, no statistical tests of Eq. (6) exist, to the best of my knowledge.

The only meaningful test is the physical requirement that the value of the correlation function be the largest for lag equal to zero (the day of the earthquake). Therefore, cross-correlations Eq. (6) were computed between $X_{\bar{\mu}}$ along normal mode periods, and $Y_S$. I used three geophysical Earth models for this: Jeffreys-Bullen "B", a 1967 model based on the compress-ibility-pressure hypothesis [16], Dziewonski-Gilbert UTD124A', a 1972 model containing sharp density discontinuities [32] [33], and Zharkov 1990 model [34]. Models were selected such that they differ significantly in the way they represent the Earth, i.e., they ought to belong to different research groups and are separated in time by at least a decade. I selected M6.3 as the optimal cut-off magnitude so as to avoid interference due to weaker earthquakes [18], and to reduce the computational load. The cut-off magnitude was lowered to M6.0 for deep (> 399 km) earthquakes so as to increase the sample size to over 50.

More than 15 000 computations of Eq. (6) between the diurnal oscillation magnitudes $^D\bar{\mu}_\omega$ (Eq. 5) obtained as daily averages from over 200 million gravity measurements, and the global >M6.3 seismicity have returned high correlation values. For the three Earth-models respectively (chronologically) the correlation values reached 0.45, 0.39, 0.39 on the day of the shallow earthquake, Fig. 2*a*, and 0.63, 0.65 and 0.67 at three days before the deep earthquake, Fig. 2*b*, as well as 0.97 in case of deep earthquakes (likely occurring along the astenosphere-mesosphere interface) at 300–400 km depths, for all three models.

In all cases (models), the strongest response was in mantle-sensitive $_0P_7$ and $_0T_7$ modes, while the lithosphere-sensitive $_0P_2$ and $_0T_2$ also returned high correlation values with deep earthquakes, Fig. 2b. Depth-separation revealed a 3-days delay in correlation, Fig. 2b. [35] speculated in a statistical study of 7359 deep ≥M3 earthquakes that a delay in deep rupture might be in effect, without proposing the duration of such a delay. Then, given that $^D\bar{\mu}_\omega$ are diurnal averages, if it exists I assume that for most earthquakes such a deep-rupture delay is either $\Delta t_1 \ll 1/2$ day, or $\Delta t_2 > 1$ day. As deduced from the variance spectra (but not the power spectra) of SG gravity, the gravity-seismicity correlation has an absolute maximum for periods ~821 s or eigenfrequencies ~1.22 mHz [3].

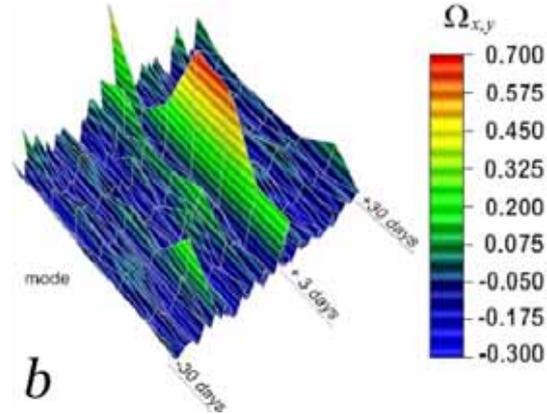

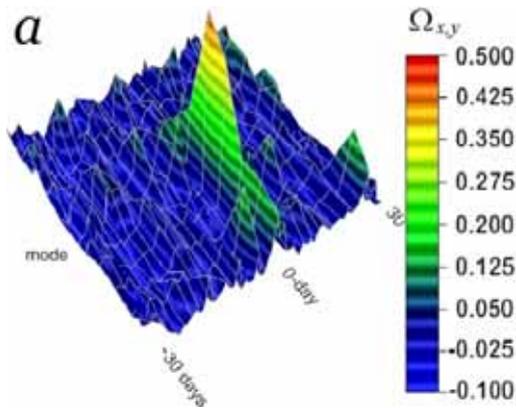

**Fig.2.** Panel *a*: time-period-correlation surface of cross-correlation ($\Omega_{x,y}$) changes between the $^D\bar{\mu}_\omega$ from 1991-2001 and seismic energies from >M6.39 earthquakes that affected Cantley SG between 1991-2001, for poloidal periods of Jeffreys-Bullen "B" model. Panel *b*: time-period-correlation surface of $\Omega_{x,y}$ changes between the $^D\bar{\mu}_\omega$ from 1991-2001 and 54 deep >M5.99 earthquakes that affected Cantley SG between 1991-2001. Time-scale lag in days. Normal-mode long periods ordered monotonically from longest (farther) to shortest (closer). Earthquakes were in seismic energies.

In all three Earth-models tested, gravity-seismicity correlations were higher when seismic energies were used rather than seismic magnitudes, as well as when variance spectra were used rather than power spectra. Also, the more recent model produced higher correlation in case of deep earthquakes.

## 4. Spectrum of Earth long-periodic oscillation

References on "tidal triggering of earthquakes" are numerous [36]. For example, a correlation was found between tidal potential and times of occurrence of earthquakes on the lunar synodic time-scale at ~14.8 days [37] [38] [39], while [40] asserted azimuth-dependent tidal triggering in regional shallow seismicity. However, many of the tidal triggering claims have been disputed. For instance, [41] dismissed [38], stating that only semidiurnal tide has sufficient power to trigger earthquakes. [42] did not even make reference to fortnightly periods in discussing tidal triggering, while [43] questioned [40] proposing that it be tested "*whether it is the oscillatory nature of tidal stress – rather than its small magnitude – that inhibits triggering*" [43]. A century of "tidal triggering" reports can be summarized as [20]: *"The following periodicities in earthquake occurrence have been proposed at one time or another: 42 min, 1 day, 14.8 days, 29.6 days, 6 months, 1 year, 11 years, and 19 years. (...) Yet a Fourier analysis of earthquake time series fails to detect significant spectral lines corresponding to luni-solar periods."*

I next spectrally analyze all the decadal $^D\bar{\mu}_\omega$ time series computed along normal modes of the Zharkov model as the most recent of the three Earth models used. Note that in the above, correlations Eq. (6) were checked by methodology, i.e., by using: (i) three geophysical Earth models, (ii) surface instead of moment seismic magnitudes, (iii) the lowermost part of the Earth natural band, (iv) variance- versus power-spectra, and (v) seismic energies versus seismic magnitudes.

In order to check the extracted periodicity of the *spectrum of the spectra* of gravity, I look at the entire computational procedure as a filter and compute its magnitude-frequency response. Of concern here is the fact that filters can enhance or reduce spectral amplitudes of certain frequencies. Response in var% of the processing viewed as a filter, i.e., of a data processing procedure based on spectral analyses should determine whether any classical noise, naturally measured by variance, was dominating the extraction of the spring-tidal periodicity from gravity spectra. To compute this response, white noise was fed to the computational procedure; meaning a "month long" test-record was processed containing random numbers between (0, 1), limits inclusive. The resulting Fig. 3 shows that the processing generally acted as a low-pass filter in the 12 days – 200 days interval. Hence, ~15 days periods most likely are not a consequence of filter amplification.

All 16 normal $^D\overline{\mu}_\omega$ series are found periodic with 14.71 days, Fig. 4, where the theoretical solar semi-annual period $S_{sa}$= 182.6211 days was enforced (suppressed). This period is in agreement with "tidal tri-ggering" reports of 14.8 days such as [38].

The maximum magnitudes on Fig. 4 are limited by the frequency-magnitude response of the processing designed to suppress the impact of classical noise on spectral magnitudes at long periods of around half month, Fig. 3. Although largely limited by the processing, the 14.71-days period still exceeds the 99% confidence level, Fig. 4, indicating a high strength of this period. All other periods shorter than 13 days and longer than 16 days, seen on Fig. 4 as not reaching the 99% confidence level, are artifacts of the processing.

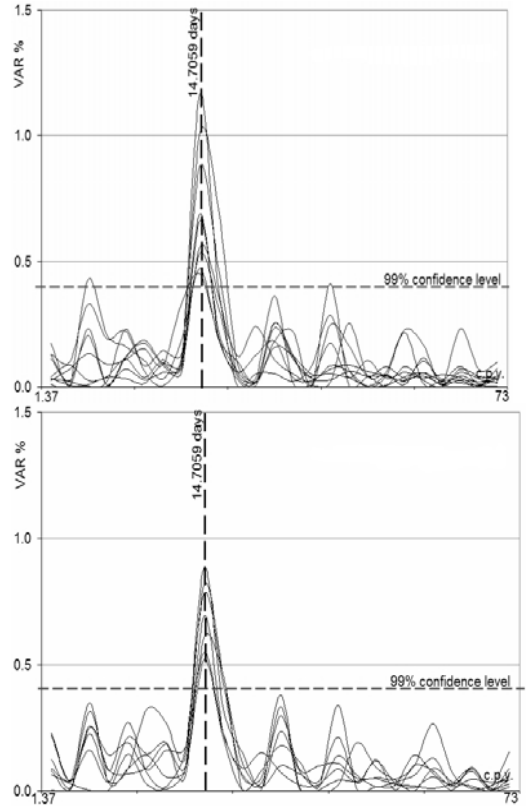

**Fig.4.** Lunar synodic periodicity of Earth's daily oscillation magnitudes, along all poloidal (panel *a*) and torsional (panel *b*) normal mode periods, Zharkov model. Enforced (suppressed) theoretical solar semi-annual $S_{sa}$= 182.6211 days.

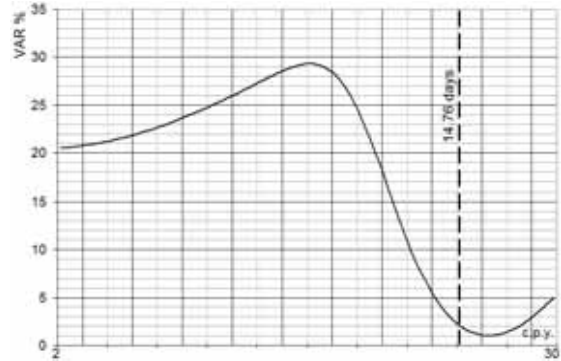

**Fig.3.** Magnitude-frequency response of the processing viewed as a filter.

If the spring tide causes the 14.71-day period, this period's estimate should get fully enhanced in the series of $^D\overline{\mu}_E$ obtained over all spectral lines from the $E_l \in$ 12–120 c.p.d. band of interest. That should improve this period's estimate to about the size of the filtering step, since the $^D\overline{\mu}_E$ series reflects the dynamics of the total-mass Earth as affected by the spring-tide. In that case, it should also be expected for the accuracy of the lunar synodic period estimate to improve even further – by simply increasing the spectral resolution.

A *lunar synodic month* of ~29.5 days is the mean interval between conjunctions of the Moon and the Sun [38]. It corresponds to one cycle of lunar phases. A more exact, empirical expression that is valid near the present epoch for one lunar synodic month is based upon the lunar theory of [44]:

$$\left.\begin{array}{l} T_{syn} \text{ [m.s.d.]} = 29.5305888531 + \\ + 2.1621 \cdot 10^{-7} \cdot T_{t.d.t.} - \\ - 3.64 \cdot 10^{-10} \, T^2_{t.d.t.}; \\ T_{t.d.t.} \text{ [JC]} = (JD - 2\,451\,545)\,/\,36\,525 \end{array}\right\} \quad (7)$$

where $T_{syn}$ is measured in *mean solar days* (m.s.d.), and $T_{t.d.t.}$ in Julian days (JD) and Julian centuries (JC) of *Terrestrial Dynamical Time* (T.D.T.) that is independent of the variable rotation of the Earth. Any particular phase cycle may vary from the mean by up to seven hours [44]. It thus sufficed for all purposes to compute spectra with a resolution better than 3½ hours locally around the period of interest. Using the 2000 pt spectral resolution enables for spectral estimates to be claimed to better than 1.3 h.

The $^D\overline{\mu}_E$ series turned out to be periodic with 14.77876 and 182.61419 days, Fig. 5. The first period is longer than the lunar astronomical fortnightly $M_f$ = 13.66079 m.s.d. by 26h 49m 53s, and also longer than the closest theoretical fortnightly tide of $M_{sf}$ = 14.76530 days by 19m 23s. Note that the latter fortnightly tide is an order of magnitude smaller in amplitude than the strongest tide $M_f$ excited by the largest tidal potential $P_2^0$ [45]. The second period extracted is the astronomical semi-annual solar, to within 9m 57s. After suppressing the theoretical solar period $S_{sa}$= 182.6211 days, the lunar period's estimate becomes 14.76053 days, at a high 26.1 var% and over 95% confidence level.

Increasing the spectral resolution to 50 000 pt enabled estimate of the latter period to better than 3 minutes, as 14.7655 days at 95% confidence level, Fig. 5. According to Eq. (7), this represents an accuracy of ~17 seconds or twice the size of the filtering step. Thus, a considerably improved estimate of the periodicity of the Earth gravity oscillation was obtained when the complete environment information is used, as well as after the spectral resolution was enhanced 25 times in this case. Note again that a distributions contradiction forbids uncertainty estimate for this value too.

The lunar synodic semi-monthly and solar semi-annual periods, as the only periods from the 1 day ($^D\overline{\mu}_E$ resolution) to 10 yr ($^D\overline{\mu}_E$ size) period interval cannot represent a residual from data processing as no systematic noise was cleared from the record to begin with. Also, the processing did not amplify these two periods as it was, in fact, designed to suppress all the contents at around half a month; see Fig. 3. Thus, the two periods are not in the SG data per sé, and the entire data, i.e., the field they sample, oscillate with those two periods. This sort of sensitivity was attainable due to SG accuracy, discussed earlier, appreciably exceeding the ratio between the force of Earth gravity and the maximum lunar tidal force, of ~1.14:10$^{-7}$ [46].

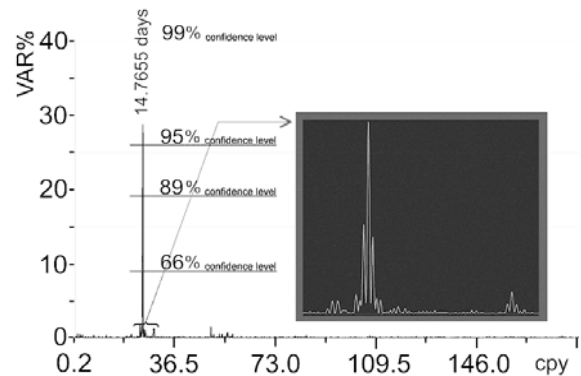

**Fig.5.** Lunar synodic half-month periodicity of Earth total mass. Shown is variance-spectrum of Earth's mean-diurnal oscillation magnitudes, obtained from variance-spectra of SG-gravity after the removal of theoretical solar semi-annual period. Data span 10.3 yr. Band 2 day to 5 yr. Spectral resolution 50 000 pt.

Obviously, since we speak of gravity, the inclusion of gravity components that are due to non-solid Earth masses (such as the atmosphere), as well as of those in the deep interior (lower mantle-to-inner core) had little importance in inflating these two periods – due to low density and distance from the lithosphere, respectively. Here, normal eigenperiods were used because they are the natural beat periods of all Earth masses that comprise gravity and cause the Earth oscillations in general, not just the Earth free oscillation. I therefore checked the periodicity obtained along all normal mode periods, Fig. 4, against the result from using complete environment information, Fig. 5. As seen above, this resulted in an excellent accuracy of the lunar period estimate, confirming the starting premise on both the SG accuracy as well as the GTS validity.

## 5. Earth as a Moon-forced mechanical oscillator, Concept of

Given the reasoning for tidal forcing of the lithosphere [47] [48] [49] [50] [51], I propose based on the above discussion and my SG observations the following

> ***Concept***
> *Earth is a viscous, stopped up mechanical oscillator forced externally mostly by the Moon's orbital period due to which the states of a maximum mass (gravity) oscillation magnification and a maximum stored potential energy occur.*

A georesonator's total mass oscillates with the springtide period, so the Earth oscillation is not just free but constrained as well; meaning rather than observing the response of the Earth as a system under exclusively free motion, as it has been done in geophysics so far, this specific concept requires that the response of the Earth system under forced motion be observed instead. Then in such a global forced oscillator first proposed by Tesla [17], where the damping force is proportional to the velocity of the body [21], the Earth grave mode $_0P_2$ is the system's normal period and the lunar synodic period is the system's forcing period due to which the states of a maximum oscillation magnification and a maximum stored potential energy occur [52].

Let us substitute in the mechanical oscillator equations [52] the $T_o$ = 3233 sec [34] grave mode as that which makes the normal $\omega_o = 2\pi / T_o$, and the SG-measured $T_{max}$ = 14.7655 days as that which makes the maximum magnification forcing frequency $\omega_{max} = 2\pi / T_{max}$. Subsequently, the spectral spread of the system response about the Earth resonant frequency $\omega = \omega_{max}$ can be obtained for the characteristic mantle viscosity of $\sim 10^{21}$ Pa s [53], as [52]:

$$Q = \frac{1}{\sqrt{2 - 2(\omega/\omega_o)^2}} = 0.70711 \cong \frac{1}{\sqrt{2}} \quad (8)$$

as well as the phase shift of the (response function of the) steady state solution of the Earth displacement, as [52]:

$$\phi(\omega_{max}) = \arctan\left(\frac{1/Q}{\omega_o/\omega - \omega/\omega_o}\right) = 0.205 \quad (9)$$

of the forcing period $\phi(\omega_{max})$ = 14.7655 days, or 3.03 days. This phase is the time by which the displacement lags behind the driving force [52]. Theoretical value Eq. (9) agrees with the observed 3-days delay in the here discovered gravity-deep seismicity correlation, Fig. 2b.

Let us now use the same periods as in the above to obtain the maximum displacement on or within the Earth due to magnification of forced oscillation of Earth masses (gravity), as [21][52]:

$$X(\omega) = \frac{F_o}{k} \cdot \left(\frac{\omega_o/\omega}{\sqrt{(\omega_o/\omega - \omega/\omega_o)^2 + 1/Q^2}}\right), \quad (10)$$

where the Earth-Moon maximum gravity force $F_{perigee}$ = 2.2194·$10^{20}$ N is the maximum forcing amplitude $F_o$, and $k = m_E \cdot (\omega_o)^2$ is the system spring constant [52]. For the Earth mass $m_E$ = 5.9736·$10^{24}$ kg (cf. NASA), and $_0P_2$ values from the three Earth models used: 3223 s, 3228 s, 3233 s, respectively, I obtain the maximal displacement Eqs. (5)–(7), as [52]:

$$x(t) = X(\omega_{max}) \cdot \cos[\omega t - \phi(\omega)]. \quad (11)$$

For shallow earthquakes, where $t = t_o = 0$, $\phi(\omega) = 0$, I obtain 9.78 m, 9.81 m, 9.84 m, respectively (chronologically). For deep (<400 km) earthquakes, where $t = t_o = 0$, $\phi(\omega) = \phi(\omega_{max})$, I obtain 9.57 m, 9.60 m, 9.63 m, respectively (chronologically). Estimates of the grave mode $_0P_2$ from the three Earth models used are seen as increasing by ~5 s per decade, probably due to the Moon receding from the Earth (tidal friction), and to totality of other factors such as the accretion of cosmic particles. It consequently appears that the maximum mass displacement on the Earth increases steadily at a rate of ~3 cm/decade. This perhaps substantiates physically the correlation between the Earth gravity and deep earthquakes, here observed as ever improving decade to decade. In case of shallow earthquakes however, the observed correlation trend is seen not as clearly. This could be due to the lithosphere's rigid-elastic environment being subject to stress-strain buildup and thus an apparent tidal inhibitting. This would also be in agreement with the report by [43], but also with a general understanding that the size of large strike-slip earthquakes (in this study the most-used type) is not related just to the amount of stress-strain buildup [54]. Earthquake triggering is a phenomenon that does not arise exclusively due to the release of static strain by the foregoing earthquake. Instead, seismicity gets activated also by direct shaking, i.e., mechanically, e.g., [55] [56]. To explain this, the Earth must be regarded as a mechanical oscillator. Finally, any stress-strain buildup is virtually absent in the mantle where a forced-oscillator model seems to be able to satisfy the observed trending of the gravity-deep seismicity correlation. Note that during the great ~M9.3 Sumatra earthquake of December 2004, numerous reports have put this displacement at ~10 m average. Also, the ~M9.5 seem to be the strongest earthquakes possible.

The spring tide exerts pull on all Earth masses. In case of the mantle's plastic environment, this pull seems to be affecting the mantle at the 300-400 km depths. The absolutely highest gravity-deep seismicity correlation was at the ~821 sec eigenperiod, at the Pacific Rim. Inserting $T_o$ = 821 sec into Eqs. (8) and (9) yields that region's own ~18.5 h phase shift. The <1-day phase shift could be due to the facts that (a) shallow earthquakes correlate with Earth oscillation magnitudes on the day of the earthquake and (b) this correlation is worse than in case of deep earthquakes.

When a mass composed mostly of the mantle or/and the crust attains its maximal resonant magnification under the Moon-Earth orbital tone, the material of which that structure is composed crumbles, resulting in an earthquake (of unspecified type). The here proposed georesonator concept also agrees with the well known fact that shallow and deep earthquakes belong to two different processes: plastic and rock environments have radically different structure and therefore considerably different structural eigenfrequencies as well. The longer the object's eigenperiods the shorter the time such an object under forced osci-llation magnification to fall apart – clearly the case of lithosphere v. mantle.

It was recently demonstrated that mantle melts take only a few decades to generate, transfer, accumulate and erupt, opposite to previous estimates of ~$10^3$ yrs [57]. The here presented georesonator concept offers an obvious mechanism for rapid transport of mantle melt, where tectonics is partially a consequence of the resonating mantle's fast dynamics at a swift rate that equals the time which takes for the 3-days phase to start affecting the SG sensitivity.

In the view of the proposed concept and the related discussion, I regard as demonstrated the following

> **Rule:**
> *The ratio of seismic energy $E_S$ and total kinetic energy $E_K$ on Earth is constant.*

Based on the above-defined concept and its *own rule* (a claim which holds only under certain conditions; a law in a non-strict sense), I propose the following extension of the proposal that large earthquakes could be caused by mantle's magnified resonance:

> **Physical hypothesis "B"**
> *Earth tectonics and inner-core differential rotations arise mainly due to respective upward and downward continuations of the Moon-forced resonant dynamics of the mantle.*

I leave the physical hypothesis "B" at the level of speculation (untested), for if it holds it would require that each tectonic structure have its own grave mode and structural eigenfrequencies as well as its own phase shift, all of which puts such testing beyond the scope of this paper.

Importantly, Eqs. (8)–(11) produce nonsensical results for the Earth–Sun system. That is expected since, unlike the Moon, the Sun "orbits" about the Earth's respective barycenter only apparently. This further legitimizes the proposed concept by providing relevance for the real Earth. Note that tidal analysis theory allows for the solar declinational tide $S_{sa}$ (as a major spectral peak) to be a mixture of gravitational and meteorological effects, and as such far from any natural resonance. Also, some claim that GPS measurements collected continuously within a 24-hour observation window reveal that diurnal peak aliasing could cause the semi-annual period to appear in the signal, but, curiously, not the lunar synodic semi-monthly period [58].

## 6. Discussion

Superconducting gravimeters can sense globally minute (in the order of 0.01 var%) mass re-distributions of the inner masses. Thus contrary to some opinions, such as [59], depending on the choice of signal and noise the removal of atmospheric and other environmental effects is not necessary for all purposes. Note that, while the gravity-seismicity correlation Eq. (6) is frequency-dependent, the relationship between local pressure (here the second-largest information constituent) and SG-sampled gravity can be independent of frequency and epoch [60]. (Of course, this is not a general rule, as the epoch dependence can indeed be expected to arise seasonally, and the frequency dependence due to local pressure albeit from much wider band than the 12-120 c.p.d. used here.) Furthermore, the air pressure- and gravity-spectra are not expected to show any resemblance [61]. Finally, the amplitude of the theoretical tidal gravity signal, observed here at 0.068 c.p.d. reaches at least 1–2 μGal [45], corresponding to a change of 5–10 mm vertically, or 1.2 –2.5 m in latitude. This is well within the Cantley SG accuracy [2], so that any signal possibly found in the spectra of SG data cannot be exclusively due to geodetic effects of either height or latitude.

As the Earth masses bulge during the springtide, their oscillatory movement attains its locally maximum amplitude during each new and full Moon. As seen in Section 4, I detected these conjunction and opposition peaks in the spectrum of the spectra of SG gravity. No intermediary that would either introduce or amplify the two periodicities could exist, since the perfect periodicity of the $^D\bar{\mu}_E$ series is superimposed naturally onto the periodicity of any information constituent alone and hence of the $^D\bar{\mu}_\omega$ series too. This voids the possibility for any intermediary to introduce the two $M_{sf}$ peaks anew. This causality, along with the above theoretical delay which I also observed experimentally as shown in Section 3, and the maximal well-known displacement which I also obtained theoretically in Section 5, constitute a sufficient condition, while the $^D\bar{\mu}_\omega$ – seismicity correlation at mantle-sensitive and lithosphere-sensitive modes constitutes a necessary condition for the Earth's lithosphere and mantle to respond to the two periods by rupturing. Rather than, as speculated by some, directly and simply "triggering" the earthquakes, the revealed periodicities in the proposed concept either trigger the Earth oscillation itself, or add to the oscillations that were already triggered by earthquakes. Then, the mechanism behind the discovered correlation can be either (*i*) the direct excitation, or (*ii*) an excitation through earthquakes that have been, in turn, triggered by long periodic tides. Based on the computations of Eqs. (8)–(11), and on the ideas of [41] and [42], I discard the latter option (*ii*) above. Then the proposed concept does not allow for the so called "tidal triggering of earthquakes" to exist as such. Furthermore, in reality, oscillatory nature of fault rupturing has long been established [62], as well as certain related phenomena apart from tides, which can trigger earthquakes [63].

Thus, regarding the Earth as a forced mechanical oscillator could perhaps help to explain the Earth tectonics. This would require that shallow earthquakes do not depend exclusively on stress-strain conditions, when actual prediction of medium-to-large earthquakes becomes possible. Such prediction would need to determine structural eigenfrequencies of separate mass bodies and systems, establish which faulting types are most sensitive to the seismicity–gravity correlations, Fig. 2, and subsequently monitor under combined Earth-Moon orbital regimes the structural eigenfrequencies of separate masses of interest, akin of structural eigenfrequencies studies of an engineering object under seismic shaking. The basic idea behind such a concept of earthquake prediction is the simplest one of all, and it resembles the infamous example of soldiers marching across a bridge.

## Corollary speculation

According to the forced global oscillator concept, all gravitational considerations performed in the vicinity of the Earth, such as those aimed at determining *G*, e.g. [1], had failed to account for lunar magnification of total-mass (gravity) oscillation during the considerations, Eq. (7), or to factor in the concerned location, $\omega_o / \omega_{max}$, as well as failed to take into account the varying $_oP_2$ of the Earth total mass. The maximum magnification is [52]:

$$M(\omega) = \frac{2Q^2}{\sqrt{4Q^2 - 1}}, \qquad (12)$$

which for the maximum values gives a scaling factor of $M(\omega_{max}) = 1+2.66 \cdot 10^{-11}$. The herein used total-mass representation of the Earth results in measuring the natural period of the Earth total mass as $T_o' = T_o + \varepsilon_T = 3445$ s ±0.35% where uncertainty is based on 1000 pt spectral resolution. This estimate is in agreement with Benioff's [64]; see Appendix.

The $2.66 \cdot 10^{-11}$ scaling factor translates into the maximum force at perigee of $5.9 \cdot 10^9$ N. Then the spring-tidal resonance can be speculated as responsible for the unmodeled portion of Earth nutation, of 10-50 milli-arc-second and presumed to be resonant-periodic with ~30 days period [65]. In the realm of astronomically forced oscillators, rescaling of gravitational force, Eqs. (7)–(12) might be necessary for all mass considerations within each tidally locked system of two or more heavenly bodies. That it would not be unusual to relate physical (mass, gravity) with geometrical (orbital periods) properties, is supported by reports of correlations relating physical and geometrical quantities, like the mass and periods of transiting planets [66].

Note that if the Earth were represented to a good approximation (such that all terrestrial determinations of *G* are consistent to within measurement precision) by a Moon-forced georesonator, the Earth would then necessarily owe (a part of) its magnetic field to the forcing oscillation [52].

Also, numerous reports of seismic precursory qualities of atmospheric data would get readily explained, as atmosphere is merely a part of total-Earth-masses resonating system. If atmosphere (gas) were viscous enough, we would be experiencing strong atmospheric seismicity too.

## Conclusion

A gravity-seismicity correlation with three-day phase is found after comparing the decadal SG gravity of total-mass Earth with the energy emitted in some 50 strong deep earthquakes from the same decade. For this, a concept of raw data as *the* criterion in evaluating a physical claim's validity was used. It was shown that the mechanical oscillator equations for a Moon-forced georesonator could successfully model such a 3-days phase. The same equations render maximal particle displacement on Earth as about 10 m, which agrees with the commonly observed displacement by gravest possible earthquakes. Based on the used geo-resonator's characteristics, this suggests that the Earth tectonics could be arising due to the springtide-induced magnification of the mantle resonance, i.e., in addition to or instead of the existing tectonogenesis hypotheses. The proposed mechanism of course cannot apply to all earthquakes or to entire tectonics. However, since (in theory) a forced-oscillatory displacement is fully describable, many if not most large earthquakes could be predictable – as a consequence of the structural collapses occurring when the magnified resonance of the mantle matches the grave mode of oscillation of the observed mass body of interest.

I speculate that mass-resonance magnification is a candidate cause for: absurdity of terrestrial $G$ experiments, inner core's differential rotations, the unexplained periodic portion of the Earth nutation, the seismic-precursory qualities of atmospheric data, and the Earth magnetic field.

# APPENDIX

In order to verify the grave mode of the Earth's total-mass oscillation I take the SG gravity recordings during three greatest earthquakes from 1990-ies that affected the Cantley SG record, Table 1. Using to GVSA unique ability to process gapped records, I look for differences between the spectra of selected gravity records without gaps and records with gaps artificially introduced. I thus make 5, 21, and then 53 filter-step-long (8 and 32 sec) gaps in the three records respectively, where the order of earthquakes was randomly selected. By observing the differences between Gauss-Vaníček (G-V) spectra of complete v. incomplete records, I look for the first instance when this difference reaches the zero value. Since both the complete and incomplete records always described the same instance and the same location when and where the same field (in this case the Earth gravity field) was sampled during the three energy emissions, it is precisely this value that marks the beginning of the Earth's natural band of oscillation. If this setup is correct, then the more gaps the record has should mean the more pronounced impact of the non-natural information onto the spectra, too. Indeed, Fig.6 shows that more gaps results in a clearer distinction between the natural and non-natural bands.

The effect of any known (including tidal) variations can be suppressed in the GVSA together with processing, where known periods in form of analytical functions or discrete data sets can be enforced on data. Thus, when GVSA is used, no preprocessing is required to strip the observations of tides. Here, nine semidiurnal and diurnal tidal periods were enforced [67]: 13.4098257, 13.9426854, 14.5057965, 15.0424341, 15.5742444, 28.4395041, 28.9840259, 29.5159428, and 29.9977268 °/hr. The vertical spectral lines superimposed on the spectra plots Fig.6 represent the normal mode periods from the Zharkov model.

Thus the grave mode (most natural period) of the Earth total mass oscillation is measured as $T_o' = 3445$ s $\pm 0.35\%$, where uncertainty is based on 1000 pt spectral resolution. This is in agreement with Benioff [64]. Note that the seismological community has been critical of the Benioff 1958 estimate, mulishly insisting on "noise" removal despite the fact that, strictly speaking, noise is a completely abstract concept, making it utterly absurd to insist on preferring either the signal or the noise in natural data.

| Earthquake event order | date | $M_S$ | Approximate location | φ | λ | d km |
|---|---|---|---|---|---|---|
| #48 | 11/08/97 | 7.9 | China | 35.07 | 87.32 | 33 |
| #79 | 03/25/98 | 8.8 | Ballenys, South of Australia | -62.88 | 149.53 | 10 |
| #80 | 11/29/98 | 8.3 | Pacific Ocean, Indonesia | 2.07 | 124.89 | 33 |

**Table A.** Three strongest shallow earthquakes in the Cantley SG record from the 1990-ies [3].

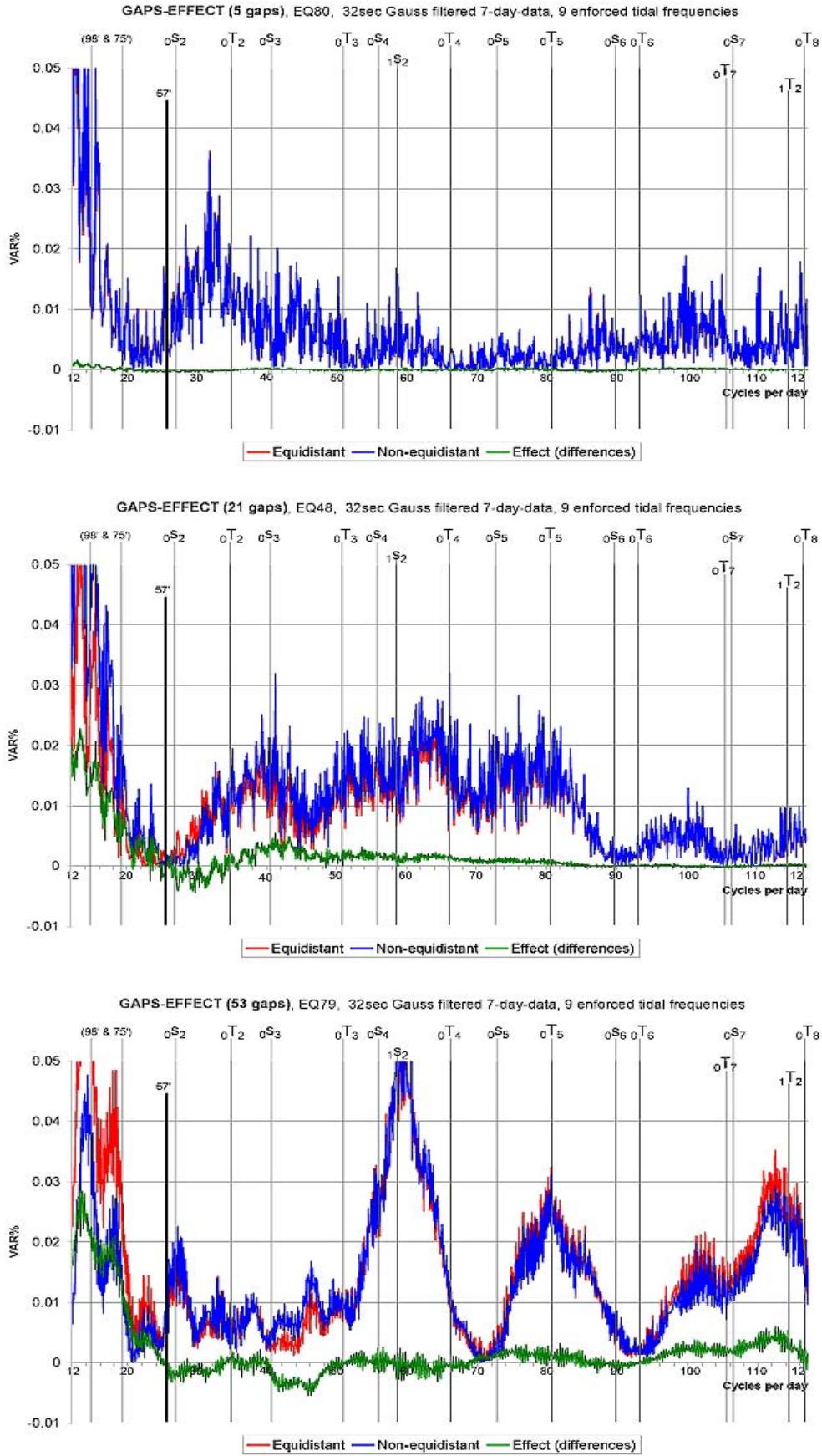

**Fig.6.** The effect (green) of regarding a series as equidistant, for 5 (top panel), 21 (mid panel), and 53 (bottom panel) 8-sec gaps. Shown are G-V spectra of gravity at one week past three strongest shallow earthquakes, Table 1.

14Omerbashich, M. (2007) *Geodinamica Acta (European Journal of Geodynamics)* 20 (6), 369-383.